\newcommand{\sgn}{{\rm sgn}}

\documentclass[showpacs,prb,preprintnumbers,amsmath,amssymb,twocolumn]{revtex4}%
\usepackage{graphicx}%
\usepackage{amsmath}%
\usepackage{amsfonts}%
\usepackage{amssymb}
\usepackage{bm}
\providecommand{\U}[1]{\protect\rule{.1in}{.1in}}

\newcommand{\etal}{{\it etal.~}}
\begin{document}
\title{Spin-orbit splittings in Si/SiGe quantum wells}
\author{M. Prada$^{1}$, G. Klimeck$^{2}$, and R. Joynt$^{1}$}
\affiliation{$^{1}$Department of Physics, University of Wisconsin-Madison, Wisconsin 53706 (USA)}
\affiliation{$^{2}$Network for Computational Nanotechnology, Purdue University, W. Lafayette, Indiana  (USA)}

\begin{abstract}
We present a calculation of the wavevector-dependent subband level splitting from
spin-orbit coupling in Si/SiGe quantum wells. \ We first use the effective-mass 
approach, where the splittings are parameterized by separating
contributions from the Rashba and Dresselhaus terms. \ We then determine the
parameters by fitting tight-binding numerical results obtained using the quantitative
nanoelectronic modeling tool, NEMO-3D.
\ We describe the relevant parameters as a function of applied electric field and 
well width in our numerical simulations. 
\ For a silicon membrane, we find the bulk Rashba parameter to be linear in field,  
$\alpha = \alpha^1E_z$ with $\alpha^1 \simeq 2\times$10$^{-5}$nm$^{-2}$. \ 
The dominant contribution to the spin-orbit splitting is from Dresselhaus-type
terms, and the magnitude for a typical flat SiGe/Si/SiGe quantum well
can be as high as 1$\mu$eV. \ 
\end{abstract}

\maketitle


\section{I. Introduction}

Silicon is a leading candidate material for spin-based quantum information
processing \cite{tahan1,tahan}. \ Its spin-orbit coupling (SOC) is relatively
weak and the hyperfine coupling can be eliminated by isotopic purification 
\cite{pradaPE}.
\ This means that the spin lifetimes should be long. \ One way to measure a
spin lifetime is to use electron spin resonance (ESR) on two-dimensional
electron gases (2DEGs) in silicon quantum wells (QWs) \cite{wilamowski, 
tyryshkin,truitt}, where the D'yakonov-Perel' \cite{dp} mechanism accounts for
the relaxation. \ SOC may also be measured directly, using 
EST \cite{wilamowski2,ensslin} or photocurrents 
\cite{ganichev}. \ In order to compare with experiment,
however, we need the wavevector-dependent SOC Hamiltonian,
which must be calculated atomistically. \ That is the aim of this paper. \ We
shall focus on Si layers grown in the $[001]$ direction with Si$_{x}$%
Ge$_{1-x}$ layers on either side.

Most calculations of the DP relaxation have used the Rashba Hamiltonian 
\cite{rashba}, which is of the form%
\begin{equation}
\label{eqrashba}
H_{R}=\alpha\left(  E_{z},N,v\right)  \left(  \sigma_{x}k_{y}-\sigma
_{y}k_{x}\right)  .
\end{equation}
$\sigma_{i}$ are the Pauli matrices and $k_{i}$ are the two-dimensional
wavevector components. \ $N$ is the number of atomic $\operatorname{Si}$
layers in the well. \ $v$ is the valley degree of freedom. \ We focus on the
lowest electric subband, in which case the valley degree of freedom is
two-valued \cite{herring} and $v$ is a two-by-two matrix.

$\alpha\left(  E_{z}\right)  $ depends (in lowest order) linearly on
$E_{z},$ the external electric field, and often only this term is kept.
\ $E_{z}$ is of order $1-5\times10^{7}$ V/m in heterostructures or 
MODFET devices \cite{charlie}. \ The large
magnitude of the field makes it important to examine the linearity assumption,
and that is one of the purposes of this paper. \ The Rashba is generally
thought of as a bulk effect. \ However, de Andrade e Silva \textit{et al}.
pointed out that surface effects may also be important \cite{andrade}, and
this assumption should also be re-evaluated. \ In addition, when a detailed
treatment of the surface effects was done by Nestoklon \textit{et al. }in the
absence of an applied field $E_{z}$\textit{, }they showed that in this case
one obtains a term%

\begin{equation}
H_{D}=\beta\left(  E_{z},N,v\right)  \left(  \sigma_{x}k_{x}-\sigma
_{y}k_{y}\right)
\label{eqdress}
\end{equation}
if $N$ is odd \cite{nestoklon,wrinkler}. \ This is a Dresselhaus-like term
\cite{dresselhaus} in that it
arises from inversion asymmetry. \ The lowest-order term in $E_{z}$ for
$\beta\left(  E_{z},N,v\right)  $ is a constant, so that if $N$ is odd
then we get a surface-induced spin splitting even in the absence of an
external field. \ 

The symmetry considerations refer to an ideal free-standing Si layer, or a
layer that is sandwiched between two identical Si$_{x}$Ge$_{1-x}$ layers that
are treated in the virtual crystal approximation. \ Real Si$_{x}$Ge$_{1-x}$
layers have substitutional disorder that destroys all symmetries.

The purpose of this paper is to determine the functions $\alpha\left(
E_{z},N,v\right)  $ and $\beta\left(  E_{z},N,v\right)  $ both for the
ideal, free-standing case and for the case of a well confined in a Si/Si$_{x}
$Ge$_{1-x}$ heterostructure. \ These functions determine the spin properties
of electrons in $\operatorname{Si}$ quantum wells. \ We shall focus
particularly on the question of which term dominates for quantum wells with
realistic values of $E_{z}$ and $N.$

The numerical tight-binding calculations are performed with NEMO-3D \cite{nemo} 
on nanoHUB.org computational resources \cite{nanohub}.
In NEMO-3D, atoms are represented explicitly in the {\it{sp$^3$d$^5$s$^*$} }
tight-binding model, and the valence force field (VFF) method is employed to 
minimize strain \cite{strain}. NEMO-3D enables the calculation of localized 
states on a QW and their in-plane dispersion relation with a very high degree
of precision, allowing to extract the splittings along the in-plane directions 
in $k$ space. \ Note that earlier calculations of $\alpha$ and $\beta$ have used the
virtual-crystal approximation for the interfaces. \ Since both of these
quantities are dominated by atomic-scale interface effects, this is a rather
crude approach.

\ This paper is organized as follows: 
Sec. \ref{secsym} discuss the symmetry operations of a silicon membrane. \ 
In Sec. \ref{valleytheor}, expressions for the SOC in 
a $\delta$-functional effective mass approach are given. \  
We present a qualitative picture in Sec. \ref{secqp}. \  
Sec. \ref{secr1} contains the numerical results for 
ideal Si QWs and  
Sec. \ref{secr2}, the ones for a SiGe/Si/SiGe heterostructure. \ 
We conclude in Sec. \ref{seccon} with a summary of the results obtained. 

\section{Symmetry \ }
\label{secsym}

We give here for clarity the symmetry operations of this system in the ideal
case, since they are far from obvious. \ We stress that this is only an
expanded discussion of the analysis already given by Nestoklon 
\etal \cite{nestoklon}. \ The
lattice considered as a bulk sample has the diamond structure with a
tetragonal distortion due to the Si$_{x}$Ge$_{1-x}$ layers: the \ [001] axis
along the growth direction is compressed relative to the in-plane [100] and
[010] axes. \ In a symmorphic lattice this would simply reduce the point group
symmetry from the cubic group $O_{h},$ with 48 operations, to the tetragonal
group $D_{4h},$ with 16. \ (Recall that a symmorphic space group is one that
is generated by translations and by rotations and reflections about a point.
\ A nonsymmorphic lattice requires combined operations such as screw axis and
glide plane operations in its generating set.) \ For the diamond lattice, a
nonsymmorphic lattice, there is no point group, but the factor group still has
16 operations, realized as follows. \ There are 4 proper rotations through
180$^{\circ}$ using the 2-fold $[100],[010],$ and [001] axes passing through
the origin, which may be taken at the position of any atom. \ In addition
there are two $\pm$90$^{\circ}$ improper rotations about the [001] axis: the
rotation is followed by a reflection in the $\left(  001\right)  $ plane.
\ There are also reflections with respect to the (110) and (1$\overline{1}$0)
planes. \ \ Any of these 8 operations may be combined with an inversion
through a point midway between any pair of nearest neighbors. \ When we
consider a layer, all 16 operations might appear to preserve the positions of
the interfaces, whose presence would therefore not reduce the symmetry.
\ However, this turns out to be not quite the case.

For a layer with an odd number $N$ of atomic layers, take the origin $(0,0,0)
$ at an atom in the central plane. \ Then the other atoms in this plane lie at
the points $\left(  a_{x}/2\right)  [n_{1}\left(  1,1,0\right)  +n_{2}\left(
1,-1,0\right)  ],$ where $a_{x}$ is the in-plane lattice constant and $n_{1}$
and $n_{2}$ denote any integer. \ Each atom in the central plane is
accompanied by another one in the plane $z=a_{z}/4$ shifted from it by
$\left(  a_{x}/4,a_{x}/4,a_{z}/4\right)  ,$ where $a_{z}$ is the lattice
constant in the $z$-direction. \ The atoms in the plane at $z=-a_{z}/4$ are
shifted with respect to the atoms in the central plane by $\left(
a_{x}/4,-a_{x}/4,-a_{z}/4\right)  .$ \ The atoms in the planes at $z=\pm
a_{z}/2$ are shifted with respect to the atoms in the central plane by
$\left(  a_{x}/2,0,\pm a_{z}/2\right)  .$ \ The positions of other atoms can
be found by translating this layer of thickness $a_{z}$ by integer multiples
of $\left(  0,0,a_{z}\right)  ,$ so to understand the symmetry properties of
the full layer it is sufficient to consider a layer with $N=5.$ \ Simple
reflection in the $z=0$ plane is not a symmetry, since it interchanges the
$z=a_{z}/4$ and $z=-a_{z}/4$ layers, whose in-plane shift is $\left(
a_{x}/2\right)  \left(  0,1,0\right)  .$ \ The $z=0$ plane is also not a glide
plane, since following the reflection by the translation of $\left(
a_{x}/2\right)  \left(  0,1,0\right)  $ to restore the $z=a_{z}/4$ and
$z=-a_{z}/4$ layers would change the $z=\pm a_{z}/2$ layers. \ The 6 rotations
using $x$-, $y$-, and $z$-axes passing through the origin are readily seen to
be symmetries of the layer, as are the reflections through the $(110)$ and
$\left(  1\overline{1}0\right)  $ planes. \ They take the point $\left(
x,y,z\right)  $ into the points $\left\{  \left(  x,y,z\right)  ,\left(
x,-y,-z\right)  ,\left(  -x,y,-z\right)  \,,\left(  -x,-y,z\right),
\right.$ $\left.  
\left(-y,x,-z\right)  ,\left(  y,-x,-z\right)  ,\left(  y,x,z\right)  ,\left(
-y,-x,z\right)  \right\}  $ which is the group $D_{2d}.$ \ This is a true
point group, and the space group is therefore symmorphic. \ Spin-orbit effects
come from terms linearly proportional to $\sigma_{i},$ the spin
operators that transform as pseudovectors, while the electric field $\vec
{E}=\left(  0,0,E_{z}\right)  $ and in-plane momentum $\vec{k}=\left(
k_{x},k_{y},0\right)  $ transform as vectors, the same as the coordinates.
\ Our interest here is in combinations of these three quantities. \ Using the
above list of operations, we find in zeroth order in $E_{z}$ that there is one
invariant term in the Hamiltonian of the form $k_{x}\sigma_{x}-k_{y}\sigma
_{y}.$ \ In first order in $E_{z},$ there is only the Rashba term
$E_{z}\left(  k_{x}\sigma_{y}-k_{y}\sigma_{x}\right)  $ (which is of course
invariant under all isometries). \ Either of these terms can be multiplied by
any even function of $E_{z},$ which is invariant under all the operations.
\ Thus we find that when $N$ is an odd number, $\beta\left(
N,E_{z},v\right)  $ is an even function of $E_{z}$ and $\alpha\left(
E_{z},N,v\right)  $ is an odd function of $E_{z}.$ \ 

For even $N,$ take the origin $(0,0,0)$ at the center of a bond between atoms
at positions $\left(  a_{x}/8,a_{x}/8,a_{z}/8\right)  $ and $\left(
-a_{x}/8,-a_{x}/8,-a_{z}/8\right)  .$ \ The origin is then a center of
inversion. \ The rotation through $180^{\circ}$ about the $\left(
1\overline{1}0\right)  $ axis is a symmetry operation. \ The $\left(
110\right)  $ axis is a screw axis since the $180^{\circ}$ rotation about this
axis must be accompanied by a translation through $\left(  a_{x}%
/4,a_{x}/4,0\right)  $ to be a symmetry operation. \ The same is true for the
$180^{\circ}$ rotation about the $\left(  001\right)  $ axis. The 8 operations
of the factor group obtained by combining these operations take the point
$\left(  x,y,z\right)  $ into the points \{$\left(  x,y,z\right)  ,\left(
-x,-y,-z\right)  ,\left(  -y,-x,-z\right)  \,,\left(  y,x,z\right)  ,
$ $
\left(y+a_{x}/4,x+a_{x}/4,-z\right)  ,\left(  -y-a_{x}/4,-x-a_{x}/4,z\right)  ,$
$\left(  -x+a_{x}/4,-y+a_{x}/4,z\right)  ,\left(  x-a_{x}/4,y-a_{x}%
/4,-z\right)  \}.$\linebreak Modulo translations, this is isomorphic to the
group $D_{2h}.$ \ Because of the appearance of the translations, this is not a
true point group and the space group is not symmorphic. \ Its action in the
Hilbert space reduces in many cases to projective rather than faithful
representations of $D_{2h}.$ \ However, this does not affect the symmetry
analysis of the Hamiltonian. \ In zeroth order in $E_{z},$ the group does not
allow any combination of terms of the form $k_{i}\sigma_{j},$ since all such
terms change sign under inversion. \ In first order, we again have the Rashba
term $E_{z}\left(  k_{x}\sigma_{y}-k_{y}\sigma_{x}\right)  .$ \ Again,
multiplication of this expression by any even function of $E_{z}$ is
permissible. \ Thus we find that when $N$ is an even number, $\alpha
\left(  N,E_{z},v\right)  $ is an odd function of $E_{z}$ and $\beta
\left(  E_{z},N,v\right)  $ is an odd function of $E_{z}.$

\section{Valley and Spin Orbit Coupling}
\label{valleytheor}

\begin{figure}[!hbt]
\centering\includegraphics[angle=0, width = 0.45\textwidth]{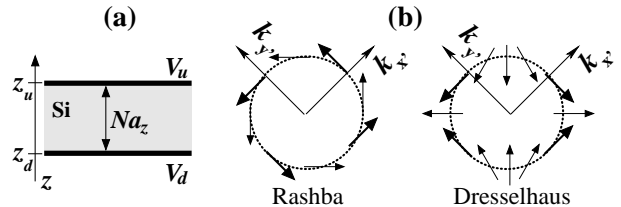}
\caption{\it \footnotesize 
(a) Rashba and Dresselhaus spin-orbit fields for
different directions in ${\bm{k}}$ space: along $\hat x^\prime$ and $\hat y^\prime$ 
directions, the eigenstates are parallel. 
(b) Schematic representation of a Si QW: the potential is considered in the interfaces 
at $z_u$ and $z_d$.}
\label{figz}\end{figure}

We take into account the above symmetry arguments and employ a 
$\delta$-functional approach for the interface-induced valley mixing 
\cite{friesenVS}. \ In the 
envelope-function picture \cite{KohnLut}, the Bloch periodic functions (valleys) at the 
interfaces $|\pm\rangle$=$u_{\pm k_{0}} e^{\pm i k_{0}L}$ mix, resulting in valley splitting, 
$\Delta_v = \Delta|\Phi(z_i)|^2\cos{(k_0L-\phi_v)}$, 
with $ \Delta$ and $\phi_v$ being phenomenological parameters, 
and $k_0$ the conduction band minima. 
$|\Phi(z_i)|^2$ is the zero-field value of the envelope function at the interfaces.
Nestoklon  {\it{et al.}} \cite{nestoklon} extend this work by introducing valley-orbit
and spin-orbit mixing, as a spin-dependent reflection of the wavefunction at the 
interfaces, situated at $z$=$z_{u,d}$ (see Fig. \ref{figz}(a)). \  
Following their approach, we introduce next $H_{\mathrm {D}}$ and 
$H_{\mathrm {R}}$ as $\delta$-functional perturbations in the lowest spinor-valleys 
functions \emph{at the interfaces}. \ We consider the hight-symmetry directions 
$x^\prime\parallel [110]$ and $y^\prime\parallel[1\bar 10]$
, along which the spin eigenstates are parallel (see Fig. \ref{figz} (b)). \ 
In this rotated basis, we have: 
$
H_{\mathrm{SO}} = 
(\alpha +\beta)\hat \sigma_{x^\prime}  k_{y^\prime}   +
(\beta-\alpha) \hat \sigma_{y^\prime}  k_{x^\prime}.
$ \ 
As mentioned in the text, a QW with $N$ even is isomorphic with the $D_{2h}$,
which contains full inversion symmetry. Under the $D_{2h}$ operations,
both $ H_{\mathrm R}$ and $ H_{\mathrm D}$ change sign:
$H_D\to -H_D$; $H_R\to -H_R$.
However, for a QW with $N$ odd, the 
point group is $D_{2d}$, which transforms: $H_D\to H_D$; $H_R\to -H_R$. 
Taking into account these symmetry arguments, we 
introduce a $\delta$-functional SOC, where the 
effective potential is at both interfaces situated in $z_u$ and $z_d$, 
as depicted in Fig. \ref{figz}(a). \ The 
advantage of the $\delta$-functional is that it allows to 
consider the symmetry arguments in a simple way, but we shall recover the 
common notation of (\ref{eqrashba}) and (\ref{eqdress}). \ The Hamiltonian 
in the $\delta$-functional approach can be expressed in terms of the 
Pauli matrices in the valley-space, $\hat s_i$: 
\begin{widetext}
\begin{equation}
\label{eq:soc_r_d}
H_{\mathrm{SO}} = \int{\mathrm{d}}\vec{r} \sum_{i}\hat s_i\left[
( \hat \sigma_{x^\prime} k_{y^\prime} - \hat \sigma_{y^\prime} k_{x^\prime})
( a_i\delta(z-z_u)-a_i^*\delta(z-z_d))
+
( \hat \sigma_{x^\prime}  k_{y^\prime} + \hat \sigma_{y^\prime} k_{x^\prime})
(b_i \delta(z-z_u)-(-1)^N b_i^*\delta(z-z_d))\right],
\end{equation}
\end{widetext}
where we have introduced $a_i$ and $b_i$ as parameters that determine the strength
of  Rashba and Dresselhaus SOC. \ We shall express those in terms of the 
commonly used $\alpha$ and $\beta$. 
Eq. (\ref{eq:soc_r_d}) can be divided in two contributions: the intra-valley SOC, 
which is Eq. (\ref{eq:soc_r_d}) with $i$ = 0, and the inter-valley SOC, 
for  $i$ = $x, y$: $H_{\mathrm{SO}}$ = $H_{\mathrm{SO1}}$ + $H_{\mathrm{SO2}}$. \

We first consider the intra-valley contribution, $H_{\mathrm{SO1}}$ ($i$ = 0). \ 
$s_0$ is then the identity matrix in the valley space, and thus,  
the spin-mixing parameters, $a_0$ and $b_0$ are real.
\ Integrating (\ref{eq:soc_r_d}) we recover the usual notation:  
\begin{eqnarray}
\label{eq:hso11}
H_{\mathrm{SO1}} &=& 
\left[ 
k_{x^\prime}\hat\sigma_{y^\prime}(\beta(E_z,N; 0)  +\alpha(E_z,N; 0))\right.\nonumber \\ &+&\left.
k_{y^\prime}\hat\sigma_{x^\prime}(\beta(E_z,N; 0)  -\alpha(E_z,N; 0))
\right]\mathbb{I},
\end{eqnarray}
where we have identified $\alpha$ and $\beta$ in terms of the
value of the envelope functions at the interfaces, $\Phi_u \equiv \Phi(z_u)$ and  
$\Phi_d \equiv \Phi(z_d)$:
\begin{eqnarray}
\alpha(E_z,N; 0) &\equiv\alpha_{0}= &a_0|(|\Phi_u|^2-|\Phi_d|^2)|; \nonumber \\
\beta(E_z,N;0)&\equiv\beta_{0}= &b_0|(|\Phi_u|^2-(-1)^N|\Phi_d|^2)|.
\nonumber\\
\label{defs1}
\end{eqnarray}
Note that the parity is contained in the Dresselhaus term, which would be non-zero for
$N$ odd and $E_z=0$ (or $|\Phi(z_u)|^2$=$|\Phi(z_d)|^2)|$). \
The Hamiltonian of Eq. (\ref{eq:hso11})  mixes spins of the same valley, which 
we denote as $\{|+\rangle\}$  and $\{|-\rangle\}$.
Remember that in this basis, the two lowest eigenstates have even or odd parity \cite{friesenVS}, 
$\varphi_{e/o} = \left( e^{-i\phi_v/2}; \pm\eta e^{i\phi_v/2 } \right)$, with 
 $\eta = \sgn \{\cos{(k_0Na_z/4)}\}$ determining the parity of the ground state, 
and $\phi_v$ defined in \cite{friesenVS}. 
We include the spin degree of freedom to diagonalize Eq. (\ref{eq:hso11}), and
find the eigen-vectors in the basis 
$\{|\uparrow,+\rangle; |\downarrow,+\rangle;
|\uparrow,-\rangle;|\downarrow,-\rangle \}$:
\begin{equation}
\label{states_hso1}
\varphi^o_{\uparrow,\downarrow} \propto\left(
\begin{array}{c}
e^{-i(\phi_v+\phi_1)/2}\\
\pm e^{-i(\phi_v-\phi_1)/2}\\
-e^{i(\phi_v-\phi_1)/2}\\
\mp  e^{i(\phi_v+\phi_1)/2}\\
\end{array}
\right); \  
\varphi^e_{\uparrow,\downarrow} \propto\left(
\begin{array}{c}
e^{-i(\phi_v+\phi_1)/2}\\
\pm e^{-i(\phi_v-\phi_1)/2}\\
 e^{i(\phi_v-\phi_1)/2}\\
\pm  e^{i(\phi_v+\phi_1)/2}\\
\end{array}
\right),
\end{equation}
with 
$\phi_1 (k_\pm) = \arg{\{ (1\pm i))\sgn(\beta_0\mp\alpha_0)\}}.$ 
\ Note that  
$\langle \varphi^i_{\uparrow,\downarrow}| \hat\sigma_{x^\prime} |
\varphi^i_{\uparrow,\downarrow}\rangle = \pm$1, 0 and
$\langle \varphi^i_{\uparrow,\downarrow} |\hat\sigma_{y^\prime}| 
\varphi^i_{\uparrow,\downarrow}\rangle =$0, $\pm 1$, so the eigen-vectors
are eigenstates of $ \hat\sigma_{x^\prime} $ or to $ \hat\sigma_{y^\prime}$, depending on the 
phase $\phi_1$ given by the direction of the in-plane wave vector, $\vec k$. \ 
Along the $\hat x^\prime$ or $\hat y^\prime$ directions, 
the spin `up' and `down' states split in both valleys by the same amount, 
$|\varepsilon_\uparrow -\varepsilon_\downarrow|=\Delta_1 \propto 2|k||(\beta_0 \pm \alpha_0)|$. \ 

We consider next the valley-mixing SOC, $H_{\mathrm{SO2}}$, which is 
Eq. (\ref{eq:soc_r_d}) with $i$ = $x,y$. \  
The parameters, $b_x$=$b_y$=$b$=$|b|e^{i\phi_\beta}$ 
and $a_x$=$a_y$=$a$=$|a|e^{i\phi_\alpha}$ ($a_z$=$b_z$=0)
are now complex, as they mix different valleys, and determine
the valley-spin coupling strength. \  
As previously noted by Nestoklon {\it et al.}, a Si QW possesses %
mirror rotation operation ${\mathcal{S}}_4$, resulting in a 
relative phase change of $\phi_{\alpha}$ ($\phi_{\beta}$) for the Rashba (Dresselhaus) interaction 
at either interface. \ Combined with the absence (existence) of inversion center for 
even (odd) $N$, a change of sign is also observed in the Rashba (both Rashba and Dresselhaus) terms. 
\ Taking into account these symmetry arguments, the valley-mixing Hamiltonian has the shape:
\begin{eqnarray}
H_{\mathrm{SO2}} = \int\left[
( \hat \sigma_{x^\prime} k_{y^\prime} - \hat \sigma_{y^\prime} k_{x^\prime})
(a \delta(z-z_u)-a^*\delta(z-z_d))
\right.\nonumber\\ \left.+
( \hat \sigma_{x^\prime} k_{y^\prime} + \hat \sigma_{y^\prime} k_{x^\prime})
( b \delta(z-z_u)-(-1)^N b^*\delta(z-z_d))\right]\mathrm{d}\vec{r}.
\nonumber
\end{eqnarray}
The valley-spin orbit Hamiltonian in the bi-spinor basis for the lowest two
valleys will thus consist of a 4$\times$4 matrix, which can be expressed in terms
of the Pauli matrices for the valleys $s_i$ and for the spin, $\sigma_i$ as: 
\begin{eqnarray}
H_{\mathrm{SO2}} &=& 
\sum_{i=x,y}  s_i \left [
\alpha(E_z,N;i) ( \hat \sigma_{x^\prime} k_{y^\prime} - \hat \sigma_{y^\prime} k_{x^\prime}) 
\right.\nonumber \\&+&\left.
\beta(E_z,N;i) ( \hat \sigma_{x^\prime} k_{y^\prime} + \hat \sigma_{y^\prime} k_{x^\prime})\right],
\label{eqso2}
\end{eqnarray}
with: 
\begin{eqnarray} 
 \alpha(E_z,N;x)&=& |a| |(|\Phi_u|^2- |\Phi_d|^2 )\cos{\phi_{0\alpha}}\nonumber \\
 \alpha(E_z,N;y)&=& |a| |(|\Phi_u|^2+ |\Phi_d|^2 )\sin{\phi_{0\alpha}} 
\nonumber \\
\beta(E_z,N;x)&=& |b||(|\Phi(z_u)|^2-(-1)^N |\Phi(z_d)|^2 )\cos{\phi_{0\beta}}\nonumber\\
\beta(E_z,N;y)&=& |b||(|\Phi(z_u)|^2+(-1)^N |\Phi(z_d)|^2 )\sin{\phi_{0\beta}},\nonumber
\end{eqnarray}
where $\phi_0=k_0L$ and  $\phi_{0i}=\phi_0-\phi_i$.  
We note that translation of the vector $\bm r$ by a three-dimensional Bravais-lattice vector 
$\bm{a}$ results in multiplication of the Bloch functions $|\pm\rangle$ by the factors 
$\exp{(\pm i k_0L)}$ \cite{nestoklon2}, and thus a phase $\phi_0$=$k_0L$ appears in the valley-mixing terms. \ 
\begin{equation}
H_{\mathrm{SO2}}
= \frac{s_z}{2} \left[\hat\sigma_{y^\prime}k_{x^\prime} (\beta_z-\alpha_z) + 
\hat\sigma_{x^\prime}k_{y^\prime} (\beta_z+\alpha_z)\right], 
\label{eq:soc22}
\end{equation}
with 
\begin{eqnarray} 
 \alpha_z
&=&|a|(|\Phi_u|^2\cos{\phi_{\alpha}^-}- |\Phi_d|^2 \cos{\phi_{\alpha}^+})\nonumber \\
 \beta_z
&=&|b|(|\Phi_u|^2\cos{\phi_{\beta}^-}-(-1)^{N} |\Phi_d|^2 \cos{\phi_{\beta}^+}),\nonumber
\end{eqnarray}
where we have defined: $\phi_{i}^\pm = \phi_{0i} \pm \phi_v$. 
Hence, we can write the SOC Hamiltonian in a compact way, 
merging Eq. (\ref{eq:hso11}) and Eq. (\ref{eq:soc22}):
\begin{eqnarray}
H_{\mathrm{SO}} &=&
\hat\sigma_{y^\prime} k_{x^\prime} 
\sum_n\sum_{i=0,z}\hat s_i \left[ \beta^{(n)}_i(E_z,N) - \alpha^{(n)}_i(E_z,N) \right]
\nonumber \\ &+&
\hat\sigma_{x^\prime} k_{y^\prime} 
\sum_n\sum_{i=0,z}\hat s_i \left[ \beta^{(n)}_i(E_z,N) + \alpha^{(n)}_i(E_z,N) \right].\nonumber\\
\label{eq:final}
\end{eqnarray}
Along $k_{x^\prime}$ ($k_{y^\prime}$), the eigenvectors are eigenstates of 
$\hat\sigma_{y^\prime}$ ($\hat\sigma_{x^\prime}$). \
The inter-valley term $i=z$, has a relative change of sign for the splittings of 
the spin `up' and `down' states in either valley, 
$\Delta_2 \propto \pm 2k(\beta_z \pm \alpha_z)$,
so there is a valley-dependent spin splitting, as depicted in the inset of Fig. \ref{figbs}:
$|\varepsilon_\uparrow-\varepsilon_\downarrow|$=$|\Delta_2\pm\Delta_1|$. \
From our numerical results, we observe that in general, $|\Delta_2|>|\Delta_1|$, causing 
a reversed symmetry in the spin structure in the lowest two valleys. \
We also observe that higher order terms contribute 
to the valley-mixing SOC, as well as intra-valley SOC in the heterostructure case. \ Hence, 
we have generalized Eq. (\ref{eq:final}) by labelling the order $n$ of the interaction, 
so far considered to zero order: 
$\alpha^{(n)} \propto (k\alpha^{(0)})^2/|\varepsilon_n-\varepsilon_0| $. \ Note that the 
numerical results presented in this work correspond to 
$\alpha = \sum_n \alpha^{(n)}$ and $\beta = \sum_n \beta^{(n)}$.

\section{Qualitative Picture}
\label{secqp}

Beyond the symmetry arguments, 
we can also analyze the functions $\alpha\left(  N,E_{z}%
,v\right)  $ and $\beta\left(  E_{z},N,v\right)  $ in qualitative terms.
\ Let us first note that there are two distinct regimes for these functions
considered in the $E_{z}-N$ plane. \ In the weak-field (WF), thick well
regime, the wavefunction for the electrons in the lowest electric subband is
spread throughout the well. \ The strong-field (SF), thin-well regime is
reached when the wavefunction is confined near one interface and does not feel
the other. \ In both the ideal and sandwich cases, the potential is rather
flat in the interior of the well over a region whose extent $\sim$ $N$ and the
confinement comes from relatively sharp interfaces. \ In this case, placing
the classical turning point in the middle of the well shows that the dividing
line between the two regimes is described by%
\begin{align*}
\text{(WF)}  & \text{: \ \ \ }E_{z}N^{3}<\frac{32\hbar^{2}}{m_{l}ea_{z}^{3}%
}=1.5\times10^{11}\frac{V}{m}.\\
\text{(SF)}  & \text{: \ \ \ }E_{z}N^{3}>\frac{32\hbar^{2}}{m_{l}ea_{z}^{3}%
}=1.5\times10^{11}\frac{V}{m}.
\end{align*}
$m_{l}$ is the longitudinal mass. \ We need to consider the two sides of this
line separately.

1. $N$ dependence of $\alpha\left(  E_{z},N,v\right)  .$ \ For $\alpha$ 
the parity of $N$ is not important. \ Let us define the lowest order
term in $E_{z}$ for $\alpha\left(  E_{z},N,v\right)  $ as $\alpha
^{1}\left(  N,v\right)  E_{z}\left(  k_{x}\sigma_{y}-k_{y}\sigma_{x}\right)
. $ \ At first sight, the Rashba effect appears to be a bulk effect and
therefore we expect $\alpha^{1}\left(  N,v\right)  $ to be independent of $N.
$ \ However, the Ehrenfest theorem implies that the expectation value of
$E_{z},$ which is proportional to the mean force, must vanish for any
wavefunction bound in the $z$-direction. \ Thus in a continuum effective mass
approximation the lowest-order term must vanish even though it is allowed by
symmetry. \ Only when we put in interface effects and other atomic-scale
effects will this term emerge. \ We shall assume that the extent of the
interface in the $z$-direction is independent of $N.$ \ If this is the case,
then the probability to find the electron at the interface in the WF regime is
$\sim1/N,$ and we may expect $\alpha\left(  E_{z},N,v\right)  $ to be a
decreasing function of $N$ in the WF regime. \ In the SF regime (large $N)$
$\alpha$ becomes independent of $N$ for fixed $E_{z}$ since we only add
layers that are unoccupied.

2. $E_{z}$ dependence of $\alpha\left(  E_{z},N,v\right)  .$ \ We have
seen that at small $E_{z}$ (WF) the dependence on $E_{z}$ is linear. \ At
large $E_{z}$ for fixed $N$ (SF) the wavefunction is increasingly squeezed
onto the interface and we may expect some continued increase in $\alpha$.
Hence, we can set:  
\[
\alpha\left(  E_{z},N,v\right) = \alpha_1\left( N,v\right) \cdot E_{z}
\]

3. $N$ dependence of $\beta\left(  E_{z},N,v\right)  .$

\ \qquad(a) $N$ odd. \ This is the only case for which $\beta\left(
E_{z}=0,N,v\right)  \neq0.$ \ In the WF regime this field-independent term may
be considered as a perturbation in $1/N,$ since it is zero for even $N$ and
the adding of an additional layer to make $N$ odd is the same as adding a term
to the Hamiltonian whose matrix elements vanish as $1/N.$ \ So we expect an
initial decrease in the term as a function of $N.$ \ Again, $\alpha$ should
approach a constant at large $N\ $and fixed $E_{z}$ for the same reasons as in
1. \ 

\qquad\ (b) $N$ even. \ $\beta\left(  E_{z}=0,N,v\right)  =0.$ \ In the
WF regime the field-independent term should converge to the result for even
$N$ as $N$ increases, since they differ by terms of order $1/N.$ \ The same
holds for the SF regime.

4. $E_{z}$ dependence of $\beta\left(  E_{z},N,v\right)  .$

\ \qquad(a) $N$ odd. \ There is a constant term but no strong dependence on
$E_{z}$ in the WF regime. \ In the SF regime the wavefunction is strongly
confined to the interface. \ If we consider just a two-layer interface, there
is a very strong orthorhombic anisotropy: the [110] and [1$\overline{1} $0]
directions are different, since the nearest-neighbor bond is in one of the two
directions. \ Hence $\beta\left(  E_{z},N,v\right)  $ can be expected to
be large, so $\beta\left(  E_{z},N,v\right)  $ should increase strongly
at large $E_{z}$ with fixed $N.$

\ \ \ (b) $N$ even. \ In the WF regime the symmetry is very important and
$\beta\left(  E_{z},N,v\right)  $ is linear in $E_{z}.$ \ Again, in the
SF regime we expect to converge to the odd $N$ result.

Finally we consider the $v$ dependence. \ The valley splitting $\Delta_{v}$
vanishes in the effective-mass continuum approximation - it is due to
interface effects. \ For $E_{z}=0,$ the eigenfunctions must be even in $z$:
$\psi_{+}\left(  r\right)  =F\left(  z\right)  \phi\left(  \vec{r}\right)
\cos k_{0}z$ or odd in $z$: $\psi_{-}\left(  \vec{r}\right)  =F\left(
z\right)  \phi\left(  \vec{r}\right)  \sin k_{0}z,$ where $F\left(  z\right)
$ is an even, slowly-varying envelope function and $\phi\left(  \vec
{r}\right)  $ is an even (in $z)$ Bloch function. \ $k_{0}$ is the wavector of
the conduction-band minima. \ For a well with smooth surfaces such as we
consider here, $E_{v}$ has the order of magnitude $\sim1-10$ meV and
oscillates with thickness on the scale $\Delta N=\pi/2k_{0}a_{z}$ and is
proportional to $1/N,$ as expected for an interface effect in the WF regime.
\ In the SF regime, $E_{z}\neq0$ and the eigenstates are no longer even or
odd. \ $\Delta_{v}$ saturates for large $N$ at fixed $E_{z},$ and its overall
magnitude increases with $E_{z},$ also as expected as the wavefunction is
squeezed onto an interface.

The oscillations with $N$ arise in the following way. \ Let $V\left(
z\right)  $ be the confining potential and $V\left(  k\right)  $ its Fourier
transform: $V\left(  k\right)  =\int\exp\left(  ikz\right)  V\left(  z\right)
dz $. \ If we apply lowest-order degenerate perturbation theory for states in
the two valleys, we find $\Delta_{v}=\left\vert \int d^{3}r~F^{2}\left(
z\right)  \phi^{2}\left(  \vec{r}\right)  e^{2ik_{0}z}V\left(  z\right)
\right\vert \sim\left\vert V(2k_{0})\right\vert .$ \ As we change $N,$
$V\left(  z\right)  $ has variation on the scale $z=4Na,$ the separation
between the two interfaces. \ $\left\vert V\left(  2k_{0}\right)  \right\vert
$ then has constructive interference when $2k_{0}\times4\left(  \Delta
N\right)  a=2\pi$ or $\Delta N=\pi/4k_{0}a.$ \ This
ignores Umklapp, which will be present in the actual system, but is absent in
the tight-binding approximation. \ 
It turns out that the dependence on the valley index can be quite dramatic. \ The valley
states differ substantially right at the interface, where much of the
spin-orbit effect arises. \ By the same token, we can expect the same
oscillations with $N$ that are seen in $\Delta_{v}$ to be present in
$\Delta_{\mathrm{SO}},$ the spin-orbit energy splitting. \ 

In actual heterostructures, the interfaces are not sharp. \ Ge is substituted
for Si on randomly chosen sites, which will generally mean that the
penetration length of the wavefunctions into the barriers varies randomly in
$x$ and $y$. \ All symmetries are violated by the disorder and it no longer
makes clear sense to speak of even and odd $N.$ \ Furthermore, the free
electrons come from dopants that create an electric field in the structure,
so that the $E_{z}=0$ limit is not accessible. \ It will probably be very
difficult to observe the parity dependences that are predicted from symmetry
arguments and also the oscillations. \ However, it may be possible to observe
such effects in free-standing membranes. \ \ \

\section{Results for Ideal Case}
\label{secr1}

In this section we get the tight-binding results for the free-standing layer.
\ Our approach to determining $\alpha$ and $\beta$ will be to compute
$\Delta_{\mathrm{SO}}$ along [110] and [1$\overline{1}$0] using NEMO-3D for
free-standing layers with varying thickness $N$ and applied electric field
$E_{z}$. \ To discriminate the Rashba and Dresselhaus contributions, 
we also compute the expectation value of
$\sigma_{x^{\prime}}$ and $\sigma_{y^{\prime}},$ which determines $\phi_1$ of 
Eq. (\ref{states_hso1}) and thus, the sign of $(\beta-\alpha)$. 

\begin{figure}[!hbt]
\centering\includegraphics[angle=0, width = 0.45\textwidth]{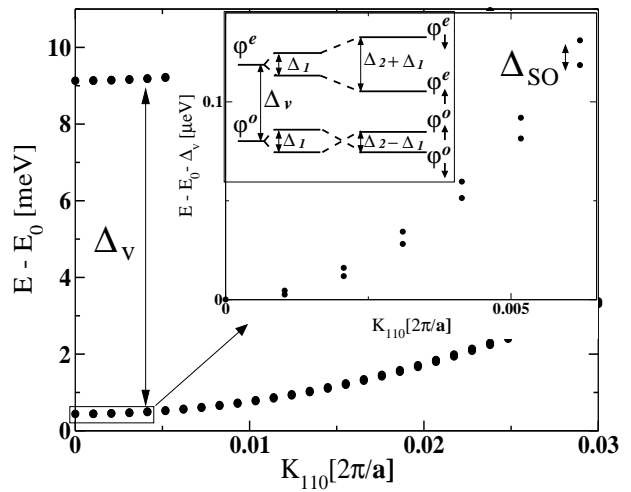}
\caption{\it \footnotesize Splitting of levels due to spin mixing within the same
valley and among different valleys. \ Note that $\Delta_{\mathrm{2}}>
\Delta_{\mathrm{1}}$, in accordance with our numerical results.}
\label{figbs}\end{figure}

In Fig. \ref{figbs} we show the four energy eigenvalues of the lowest electric subband as a
function of wavevector in the [110] direction for a 8 nm QW, 
which corresponds to $N=60$, and $E_{z}=0.$ \ The valley splitting is about 9 meV, which is much
larger than the spin-orbit splitting for all $k$. \ It is seen that\ $\Delta_{\mathrm{SO}}$ is linear in $k$ at
small $k$.
\ Extracting $\Delta_{\mathrm{2}}$ and $\Delta_{\mathrm{1}}$ (see diagram in the inset of Fig. 
\ref{figbs}) we can then calculate
 $\left(\beta+\alpha\right)  $ and $\left(  \beta-\alpha\right)  $
from the slopes of these lines, which allows us to determine $\alpha_i\left(
E_{z},N\right)  $ and $\beta_i\left(  E_{z},N\right)$ (see Eq. (\ref{eq:final})).

\begin{figure}[!hbt]
\centering\includegraphics[angle=0, width = 0.38\textwidth]{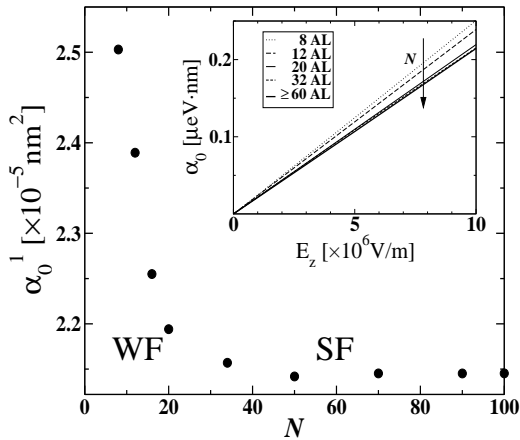}
\caption{\it \footnotesize  Rashba SOC contributions to the
splittings of the first valley subbands in Si QWs as a function of thickness, $N$, 
for a unitary electric field, $E_z$. \
Inset: $\alpha_0$ as a function of $E_z$ for different QW thicknesses.  
}
\label{figalpha}
\end{figure}

We consider first an ideal Si QW, were the symmetry arguments can be directly applied. \ 
We impose hard wall boundary conditions at both interfaces and perform tight-binding 
calculations to obtain the eigenvalues and eigenvectors. \ 
As mentioned above, we expect the Rashba contribution to be linear in $E_z$, 
$\alpha_0(E_{z},N)\approx \alpha_0^{1}E_{z}$, and   
insensitive to $N$ in the SF regime. \  
The inset of Fig. \ref{figalpha} shows that this is the case: 
The intra-valley contribution of the Rashba SOC, $\alpha_0^1$, is plotted as a 
function of $N$. \ For large $N$, $\alpha_0$ converges to a value and is insensitive
to any changes of $N$, as corresponds to the SF regime. \ 
From this data, we find $\alpha_0^{1}\simeq 2\times10^{-5}$nm$^{2}$ in the SF limit. \ 
The inset shows $\alpha_0$ as a function of $E_{z}$: for the regime of fields considered
here, it appears to be linear. \  

\begin{figure}[!hbt]
\centering\includegraphics[angle=0, width = 0.38\textwidth]{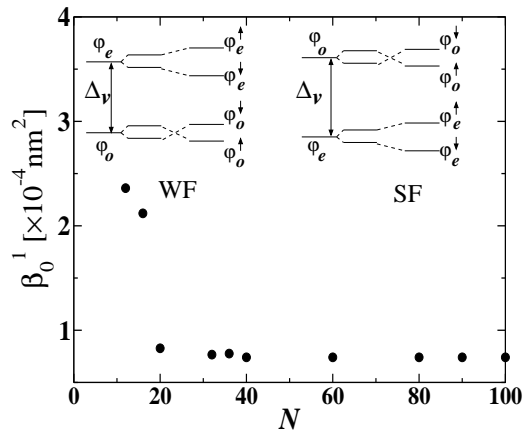}
\caption{\it \footnotesize  Intra-valley Dresselhaus SOC contributions to
splittings of the first subbands in Si QWs as a function of $N$ ($N$ even). 
The insets show the valley-spin symmetry of the four lowest conduction subbands: 
a parity flip occurs at $N\sim$ 20. 
}
\label{figbeta}
\end{figure}

For the Dresselhaus contribution, we consider first the $N$ even case. \
We do not observe any SOC related terms at $E_z=0$, as expected: 
We recall that any term of the form $k_j\sigma_i$ is violated under $D_{2h}$ 
symmetry operations. \ 
We expect the terms to be linear to lowest order of  
$E_z$, which is indeed the case:  $\beta_{0(z)}\simeq \beta_{0(z)}^1E_z$.
The results for the intra-valley are shown in Fig. \ref{figbeta}. \ 
The SF value for the intra-valley is $\beta_0^1\simeq$8$\times$10$^{-5}$nm$^{-2}$. \ 
We note an abrupt change in $\beta_0$ accompanied by a parity flip 
(depicted in the inset of Fig. \ref{figbeta}), an event that
has already been noted in literature \cite{boykin1,boykin2,boykin3,boykinVS}. \ 
This reveals that $\beta_0$ is more sensitive to higher energy contributions than $\alpha_0$. \ 

\begin{figure}[!hbt]
\centering\includegraphics[angle=0, width = .45\textwidth]{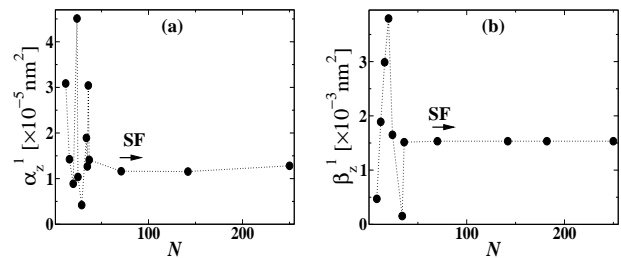}
\caption{\it \footnotesize  Linear term of the inter-valley SOC contributions to
splittings of the first subbands in pure Si QWs, 
as a function of well width. (a) Rashba and (b) Dresselhaus (for even $N$) contributions.   
The lines are a guide to the eye. 
}
\label{figab2}
\end{figure}

The inter-valley parameters are plotted next. We observe again a linear
behavior with electric field for both terms, as expected from our qualitative
arguments. \ 
Fig. \ref{figab2} shows (a) $\alpha_z^1$ ($\alpha_z=\alpha_z^1E_z$), 
and (b)  $\beta_z^1$ ($\beta_z=\beta_z^1E_z$). \
Recall that under the $D_{2h}$
operations, both the Rashba (for any $N$) and the Dresselhaus SOC (for even $N$) transform
in a similar manner, hence a similar behavior is expected with $E_z$. \ 
We observe that both
contributions exhibit oscillations as a function of $N$. \ These oscillations 
are related to the valley-splitting oscillations, 
already observed in literature  \cite{nestoklon, nestoklon2, boykinVS,friesenVS,
jancu}. \ For large $N$, the interaction converges to the SF limit value:
$\alpha_z^1\sim 1.3\times$10$^{-5}$nm$^{-2}$ and 
$\beta_z^1\sim 155\times$10$^{-5}$nm$^{-2}$. \ 

These results lead to two significant
conclusions: (i) the  Rashba inter- and intra-valley contributions are of the same order of magnitude,  
which would indicate that the bulk part of $\alpha$ predominates over the interface part in the SF
regime. \ On the contrary, $\beta_z^1$ is more than one order of magnitude larger 
than $\beta_z^0$, indicating that it is a pure interface effect, also sensitive
to higher order energy levels.  \ 
(ii) We also find from Figs. \ref{figalpha}-\ref{figab2} that the dividing line between 
the SF and WF regime is at $E_zN^3\simeq10^{11}$. 

\begin{figure}[!hbt]
\centering\includegraphics[angle=0, width = 0.45\textwidth]{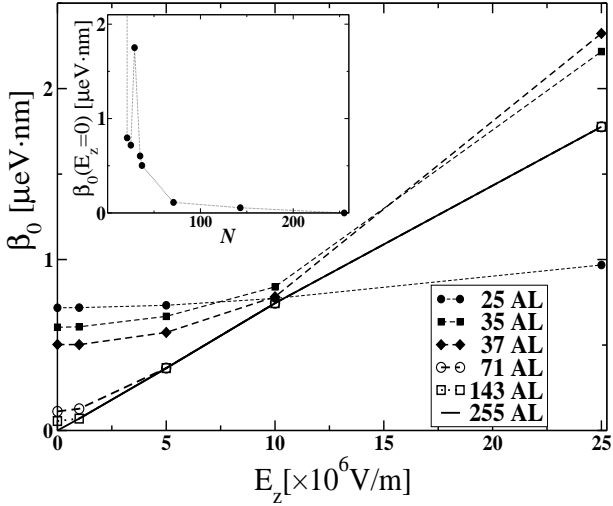}
\caption{\it \footnotesize  Intra-valley Dresselhaus SOC contributions to
splittings of the first subbands in Si QWs as a function of $E_z$
for odd $N$. \ The inset 
shows the zero-field value $\beta_0$($E_z=0$) as a function of $N$.  
}
\label{figbetaOdd}
\end{figure}

Next, we consider the Dresselhaus contribution for $N$ odd. \ 
Fig. \ref{figbetaOdd} shows $\beta_0$ as a function of $E_z$ for 
different $N$. \ We observe that for large $N$, the parity effect is less apparent,
and $\beta_0$ becomes independent of $N,$ as predicted from our qualitative arguments. \ 
In the WF regime, $\beta_z$ is not as sensitive as $\beta_0$ to electric fields. \ 
We note that $\beta_0(E_z=0)$ is non-zero, as shown in the inset of Fig. \ref{figbetaOdd}. \ 
It presents strong oscillations with $N$ in the WF regime, due to mixing with 
higher energy states, and vanishes in the SF regime, as expected in the bulk limit 
for silicon. \  
The overall $\beta_0(E_z=0)\sim1/N$ dependence is also evident in this curve.

\begin{figure}[!hbt]
\centering\includegraphics[angle=0, width = 0.45\textwidth]{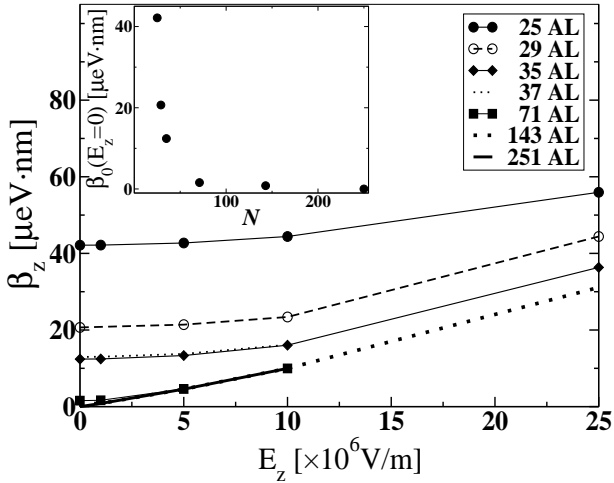}
\caption{\it \footnotesize  Inter-valley Dresselhaus SOC contributions to
splittings of the first subbands in Si QWs as a function of $E_z$, for 
$N$ odd. The inset 
shows the zero-field value $\beta_z$($E_z=0$) as a function of $N$.  
}
\label{figbeta2Odd}
\end{figure}

The inter-valley Dresselhaus coupling constant for $N$ odd is shown in Fig. 
\ref{figbeta2Odd}. \ For large $N$, $\beta_z$ becomes linear with 
$E_z$, as it corresponds to the bulk limit. \ 
The behavior is quite similar to $\beta_0$. \ 

\section{ Results for Heterostructure Case}
\label{secr2}

Having verified that the results are reasonable overall,
we redo the calculations for a more realistic model of an actual
heterostructure. \ First, we present results on a 
Si$_x$Ge$_{1-x}$/Si/Si$_x$Ge$_{1-x}$ membrane with $x$ = 0.5. 
The Si layer is surrounded by 28 layers of Si$_{x}%
$Ge$_{1-x}$ on both sides. This is sufficient to avoid surface 
effects and to confine the wavefunction in the Si QW for the electric 
fields presented here. \  Biaxial (shear) strain is minimized using VFF, which 
includes precise values of the elastic constants $c_i$, as defined elsewhere
\cite{vandervalle}. 
The unit cell is chosen to have 24 atomic layers along the $[010]$ and $[100]$ 
directions, for which realizations of the substitutional disorder are averaged 
over, as noted in our numerical data. \ These
calculations are much more time-consuming than those for the ideal layer, so
fewer results are presented. \ We recall that calculations of this kind have not
been done previously. \ Earlier work used the virtual-crystal approximation
for the outer layers, which artificially preserves the symmetry. \ The
reduction in symmetry can only increase the number of possible terms in the
Hamiltonian, so not only the Rashba and Dresselhaus terms exist, but in
principle all terms $k_{i}\sigma_{j}$ can exist - strictly speaking, even
$k_{x}$ and $k_{y}$ are no longer good quantum numbers. \ However, we shall
take advantage of the approximate symmetry to present the results in the same
way. \\

\begin{figure}[!hbt]
\centering\includegraphics[angle=0, width = 0.48\textwidth]{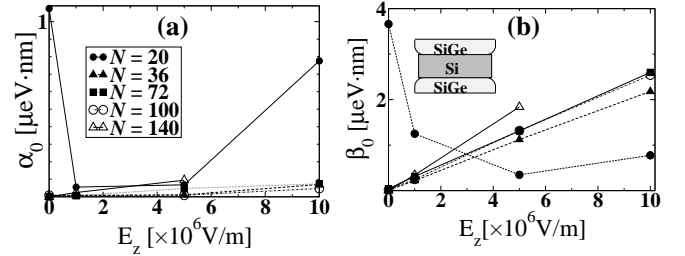}
\caption{\it \footnotesize Intra-valley contributions to SOC of a 
Si$_x$Ge$_{1-x}$/Si/Si$_x$Ge$_{1-x}$ ($x$ = 0.5) membrane as a function of $E_z$, 
(a) Rashba and (b) Dresselhaus,  for different number of atomic
layers of Si, $N$. \  The inset of (b)
sketches the structure, a SiGe/Si/SiGe suspended membrane. 
$\bullet$, $N$=20;  $\blacksquare$, $N$=36; $\blacktriangle$, 
$N$=72 ;  $\circ$, $N$=108; $\triangle$, $N$=144; 
}
\label{figalphaSiGe}
\end{figure}

Fig.  \ref{figalphaSiGe} shows the parameters $\alpha_0$ (a) and $\beta_0$ (b)
as a function of electric field, $E_z$. \  
We note that the dramatic dependence on the parity of $N$ is no longer present,
as the disorder destroys their distinct symmetry properties. We also note that $\alpha$
is always non-zero, even for $E_z=0$. \ 
The intra-valley  $\alpha_0$ is non-linear in the WF regime, 
reaching a linear-in-$E_z$ value comparable to the one of Si QW
in the SF regime (open triangles),  
$\alpha_0^1 \sim 1.9\times$10$^{-5}$nm$^2$. \ 
This is consistent with the value obtained above for the Si QW. 

Fig.  \ref{figalphaSiGe} (b) shows an overall $1/N$ dependency of  
$\beta_i$ for $E_z=0$ , consistent with the previous section results. 
For large $N$ we observe that $\beta_0$ is linear with electric fields. 
We fit QW in the 10-30nm range and find $\beta_0^1\simeq 38\times
$10$^{-5}$nm$^2$. \ We also find, for the SF regime, $\beta_z^1\simeq 58\times
$10$^{-5}$nm$^2$, with $\beta_i = \beta_i^1E_z$. \

\ We have also studied the eigenvectors for each case and find frequent parity-flips 
(see inset of Fig. \ref{figbeta}) by varying $N$ or $E_z$. As a consequence, 
both $\beta$ and  $\alpha$ depend on $N$ even in the SF regime.

Finally, we include the most experimentally relevant case: a 
Si$_x$Ge$_{1-x}$/Si/Si$_x$Ge$_{1-x}$  
grown on a Si$_{1-x}$Ge$_{x}$ substrate, with $x$ = 0.3. \ 
The in-plane lattice constant $a_\parallel$ is now relaxed to the SiGe. 
\ We apply fixed boundary conditions for the lattice
constant to the SiGe value at the bottom of the QW, and allow NEMO-3D to 
minimize the strain energy by varying the lattice constant along the $z$-axis.

\begin{figure}[!hbt]
\centering\includegraphics[angle=0, width = 0.48\textwidth]{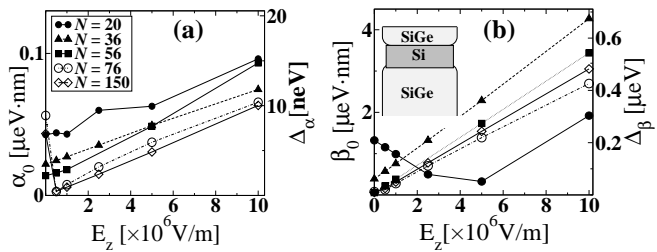}
\caption{\it \footnotesize Intra-valley contributions to SOC for a 
Si$_x$Ge$_{1-x}$/Si/Si$_x$Ge$_{1-x}$ formed on a Si$_x$Ge$_{1-x}$ substrate,
with $x$ = 0.3.  / The Rashba (a) and Dresselhaus (b) contributions are shown as a function
of $E_z$ for different number of atomic layers of Si, $N$. \ The right vertical 
axis show the absolute energy values of the splitting, calculated for the Fermi level of a
typical Si QW with $n_s = 4\times10^{11}$cm$^{-2}$ \cite{nakul}. 
The inset of (b) 
sketches the structure: a Si layer is grown in SiGe substrate and capped with a SiGe layer. 
}
\label{ab_Substrate}
\end{figure}

Fig. \ref{ab_Substrate} shows the linear-in-$\bm{k}$ SOC intra-valley 
contributions. \ The splittings are linear in $E_z$ in the SF limit, although 
parity flips and higher energy levels results in small non-linearities. \
We note, however, that the zero field shows a visible splitting, even for wide 
QWs ($N\simeq$140).  We observe that $\alpha_0$ is much smaller than in the 
previous samples, where silicon was the dominant component. 
For a typical Si/SiGe heterostructure with  $n_s = 4\times10^{11}$cm$^{-2}$ \cite{nakul} 
we have $k_F\simeq0.16$nm$^{-1}$, at which the Dresselhaus-induced SOC splitting is 
$\Delta_\beta\simeq$ 1.25 $\mu$eV, whereas the Rashba splitting is only 0.02 $\mu$eV. 

We show in table \ref{table1} the linear SOC coefficients for a 20nm Si QW ($N\simeq$140) 
in the three different structures considered in this work, for a typical electric field 
$E_z$ = 10$^7$V/m.  $\alpha_i^1$ and $\beta_i^1$ are in units of 
10$^{-5}$nm$^2$.  Note that the Dresselhaus is dominant in all cases.

\begin{table}[!hbt]
\centering
\caption{\it\footnotesize $\alpha_i^1$ and $\beta_i^1$ $i$= $0,z$ numerical
results for 20nm Si QW in the three different fashions considered in this work, 
all in units of 10$^{-5}$nm$^{2}$.}
\begin{tabular}{r r r r r }
  \hline\hline
$\alpha_i^1$($\beta_i^1$)[$\times$10$^{-5}$nm$^{2}$]
&\ \ \ \  $\alpha_0^1$\ \
& \ \ \ \ $\alpha_z^1$ \ \
& \ \  \  \ $\beta_0^1$ \ \
&\ \ \ $\beta_z^1$ \ \ \
\\\hline
`Pure' Si Membrane & 2.1	& 1.3	& 8.0	&	154.5  \\
SiGe/Si/SiGe Membrane & 
1.9 & 2.5 &  37.7 & 58.5 \\
Si QW on SiGe Substrate &
0.7 & 1.5 & 30.6 & 97.8 \\
\hline \hline
\end{tabular}
\label{table1}
\end{table}

\section{Conclusions}
\label{seccon}

We have been able to extract the anisotropic-in-$\bm{k}$ splittings due to SOC
using a $sp^3d^5s^*$ tight binding model capable to take into account
the interface effects to atomic scale. \ For a Si QW, we distinguish 
the $N$ even and odd cases, since symmetry operations over the Dresselhaus-type 
terms are fundamentally different: while for $N$ odd it appears at zero-order in $E_z$, 
for $N$ even is linear in $E_z$, to lowest order (as Rashba-type terms are). \ 
We have extracted the linear-in-$E_z$ and linear-in-$\bm{k}$ parameters for the 
$N$ even case. 

We also distinguish two regimes of operation 
for typical wells: in the weak field regime, the splittings vary strongly as a 
function of $N$. The intra-valley mixing components show roughly a $1/N$ behavior, whereas 
the valley-mixing ones present also oscillations. \ 
On the contrary, the splittings do not change with $N$ in the strong field limit. \ In the case of 
Dresselhaus, the dependency on $N$ is more prominent. \ Together with some oscillations
observed for the intra-valley mixing, this would reveal that higher-energy states
are also very important. \ 
We also observe a reverse spin structure in the spin-split valleys, a direct 
consequence of the inter-valley splitting being larger than the intra-valley. 

We find consistent results for Si QWs formed in SiGe heterostructures. \ 
We have also studied the lowest subband eigenstates, and found frequent parity flips by 
varying $N$ or $E_z$, suggesting a sample-dependent SOC. \ 
In accordance with ref. \cite{nestoklon}, we find that the energy splittings due to
Dresselhaus are in general larger than Rashba type ones: the Dresselhaus parameter 
$\beta$ is almost one order of magnitude larger than the Rashba $\alpha$ 
for the ideal (pure Si QW) case, and roughly two orders of magnitude larger 
for the heterostructure (Si on SiGe substrate) case. \ 
New experimental data fits show indeed that the Dresselhaus term is larger than the 
Rashba one \cite{data09}. \
We recall, however, that throughout this paper the numerical data have been 
obtained for QWs with flat interfaces along a main crystallographic axis. \ 
More realistic samples include a small tilted angle with respect to a high symmetry 
axis. \ The numerical data are beyond the scope of this publication, however, 
we expect the interface induced Dresselhaus parameter $\beta$ to become smaller in 
\emph{tilted} samples in a similar manner as valley splitting does \cite{kharche}. \ 
We expect, nevertheless, that $\alpha$ would remain of the same order of magnitude 
even in realistic tilted samples. \ Simulations carried out on a thin membrane 
confirm this hypothesis, although a more extensive study would quantitatively 
determine the SOC parameters. \

\begin{acknowledgments}{We would like to thank N. Kharche, M. Friesen and M. Eriksson
 for useful conversations. \ This work was supported by the Spanish
Ministry of Education and Science (MEC).
Financial support was provided by the National Science Foundation, Grant Nos. NSF-ECS-0524253 
and NSF-FRG-0805045, and by ARO and LPS, Grant No. W911NF-08-1-0482}.  
\end{acknowledgments}
\bibliography{soc_prada_09}

\end{document}